\documentclass[nohyper,notoc]{JHEP3}

\usepackage{amsmath,euscript,array,amssymb,cite,bm} 
\usepackage{graphicx}
\usepackage{subfigure}
\usepackage{axodraw}
\def\M{{{\cal M}}}

\def\Tr{{\rm Tr}}
\def\sst{\scriptscriptstyle}
\def\det{{\rm det}}

\def\Dbarslash{\,\,{\raise.15ex\hbox{/}\mkern-12mu {\bar\D}}}
\def\Dslash{\,\,{\raise.15ex\hbox{/}\mkern-12mu \D}}
\def\delslash{\,\,{\raise.15ex\hbox{/}\mkern-9mu \partial}}
\def\delbarslash{\,\,{\raise.15ex\hbox{/}\mkern-9mu {\bar\partial}}}

\let\vev=\Vev

\def\D{{\cal D}}
\def\Dbarslash{\,\,{\raise.15ex\hbox{/}\mkern-12mu {\bar\D}}}
\def\delslash{\,\,{\raise.15ex\hbox{/}\mkern-9mu \partial}}
\def\Dslash{\,\,{\raise.15ex\hbox{/}\mkern-12mu \D}}

\def\={\, =\, }
\def\+{\, +\, }
\def\-{\, -\, }

\newcommand{\be}{\begin{equation}}
\newcommand{\ee}{\end{equation}}
\def\bea{\begin{eqnarray}}
\def\eea{\end{eqnarray}}

\def\Nf{{N_{f}}}
\def\Nc{{N_{c}}}
\def\p{\varphi}
\def\pt{\tilde\varphi}
\def\uno{\mbox{1 \kern-.59em {\rm l}}}
\def\vplus{|{\rm vac}\rangle_+}
\def\vzero{|{\rm vac}\rangle_0}

\def\tb{\tan\beta}
\def\mus{\mu_\mathrm{\sst MSSM}}
\def\qsusy{Q_\mathrm{\sst SUSY}}
\def\qmess{Q_\mathrm{\sst Mess}}
\def\tten{\scriptsize $\times 10$}

\title{Patterns of Gauge Mediation in Metastable SUSY Breaking}

\author{Steven A.~Abel$^{a}$, Callum Durnford$^{a}$, Joerg Jaeckel$^{a,b}$ and Valentin V.~Khoze$^{a}$\\

${}^{a}$Institute for Particle Physics Phenomenology, University of Durham,
Durham, DH1 3LE, UK\\
${}^{b}$ Institute for Theoretical Physics, Heidelberg University,\\
${}$ Philosophenweg 16, D-69120, Heidelberg, Germany

{\tt s.a.abel@durham.ac.uk,}\,
{\tt callum.durnford@durham.ac.uk,}\,
{\tt joerg.jaeckel@durham.ac.uk,}\,
{\tt valya.khoze@durham.ac.uk}}

\abstract{
Supersymmetry breaking in a metastable vacuum allows one to build simple and concrete models of gauge mediation.
Generation of gaugino masses requires that $R$-symmetry be broken in this vacuum. In general, there are two possible
ways to break $R$-symmetry, explicitly or spontaneously.
We find that the MSSM phenomenology depends
crucially on how this breaking occurs in the Hidden Sector. Explicit $R$-symmetry
breaking models can lead to fairly standard gauge mediation, but we argue that in the
context of ISS-type models this only makes sense if $B=0$ at the mediation scale,
which leads to high $\tan\beta$. If on the other hand,
$R$-symmetry is broken spontaneously,
then $R$-symmetry violating soft terms tend to be suppressed
with respect to $R$-symmetry preserving ones, and one is led to a scenario with
large scalar masses. These models interpolate between standard gauge mediation and split SUSY models.
We provide benchmark points for the two scenarios.
They demonstrate that the
specific dynamics of the Hidden Sector -- the underlying nature
of supersymmetry and $R$-symmetry breaking --
affects considerably the mass spectrum of the MSSM, and vice versa.
}

\preprint{IPPP/07/97\\
DCPT/07/194
}

\begin{document}

\section{Introduction}

Supersymmetry continues to be the most compelling candidate for the theoretical framework
describing particle physics beyond the Standard Model. In the current paradigm, SUSY is dynamically broken in the
Hidden Sector of the full theory and the effects of this SUSY-breaking are mediated to the Visible Sector (MSSM)
by so-called messenger fields. In the usual formulation, one
essentially ignores the Hidden Sector theory and subsumes its details into
a few parameters; the scale $M_{\sst SUSY}$ at which SUSY is broken
in the Hidden Sector, the nature of mediation (gravity, gauge, extra dimensions, etc.) and the types of messenger fields.
Thus it is tempting to assume that the details of the Hidden Sector are
largely irrelevant to Visible Sector phenomenology,
and that the entire pattern of the SUSY-breaking soft terms in the MSSM is generated and determined by the messengers.
The recent breakthrough made by
Intriligator, Seiberg and Shih (ISS) \cite{ISS} in realising the dynamical SUSY breaking (DSB)
via metastable vacua, provides a very minimal and simple class of candidates for the Hidden sector,
and  makes it natural to reexamine this assumption. In particular, is it possible to distinguish different
types of Hidden Sector physics for a given type of mediation and messenger?

We shall address this question in the context of models of low scale i.e. gauge mediation (GMSB).
These were introduced in the early days of SUSY model building in
Refs.~\cite{GM1,GM2,GM3,GM4,GM5,GM6} and were subsequently revived in \cite{GM7,GM8,GM9} (see \cite{GR}
for a comprehensive review of GMSB patterns and phenomenology).
The main advantage of gauge mediation from a phenomenological point of view
is the automatic disposal of the flavour problem which plagues
gravity mediation. In GMSB the messenger fields interact only with the gauge field supermultiplets in the MSSM
and the gauge interactions do not generate unwanted flavour changing soft terms in the MSSM. The sfermion
soft masses are universal
in flavour space
and the only source of flavour violation is through the
Yukawa matrices, which is already incorporated correctly into the Standard Model.
Furthermore, the SUSY scale
in GMSB is relatively low, $M_{\sst SUSY} \ll \sqrt{m_W M_{\rm Pl}},$, and  one can determine the
full field theory in its entirety without appealing to the uncalculable
details of an underlying supergravity theory, as one has to in gravity mediation.
Indeed, the recent realisation \cite{ISS} that the dynamical breaking of supersymmetry can be achieved easily in
ordinary SQCD-like gauge theories implies that now one can formulate complete
and calculable models of gauge mediated SUSY-breaking including the Hidden (and Visible) sectors.
The goal of this paper is to study and classify these models, and to show how the generic patterns of
SUSY breaking generated in the MSSM depend on the details of the Hidden Sector.

To anticipate our findings, Visible Sector phenomenology depends
essentially on how $R$-symmetry is broken in the Hidden Sector.
Explicit $R$-symmetry breaking models of Refs.~\cite{MN,AS} can
lead to fairly standard gauge mediation, but we argue that in the
context of ISS-type models this only makes sense if $B=0$ at the
mediation scale, which leads to high $\tan\beta$. If, on the other
hand,  $R$-symmetry is broken spontaneously, as in the model of
Ref.~\cite{ADJK}, then $R$-symmetry violating operators in the
MSSM sector (e.g. gaugino masses) tend to be suppressed with
respect to $R$-symmetry preserving ones (e.g. scalar masses), and
one is led to a scenario with large scalar masses (and of course
more fine-tuning).\footnote{This kind of spectrum
indicates a residual approximate $R$-symmetry in the model as the main cause for the suppression of gaugino masses relative
to scalars. The precise reason for this suppression (investigated in a subsequent paper \cite{AJKM}) is a little more complicated
and is linked to the \emph{direct} nature of gauge mediation.}
In the limit of small $R$-symmetry breaking
we recover so-called split SUSY models \cite{split1,split2}. We provide benchmark points for the two
scenarios as an illustration.

\subsection{On the unavoidability of metastability and on $R$-symmetry breaking}

If SUSY is discovered at the LHC and is of the gauge mediation type,
then metastability of the vacuum is likely to be unavoidable \cite{ISS2}
because of two pieces of evidence:  gauginos are massive, and so too are
$R$-axions. We will briefly discuss why this is so, in full generality and independently of the models of ISS. To see that
metastability follows from these two pieces of evidence,
the first important observation is the theorem
by Nelson and Seiberg \cite{NS}, that an exact $R$-symmetry is \emph{necessary
and sufficient} to break SUSY in a generic calculable theory (of the Hidden sector).
At the same time, Majorana masses of gauginos
have non-vanishing $R$-charges.
Thus we have a phenomenological problem which
could be called
the \emph{gaugino mass problem}: gaugino masses require both supersymmetry and $R$-symmetry
breaking, but Ref.~\cite{NS} tells us that these two requirements are
mutually exclusive. How to get around it?

One approach \cite{ISS2} is to assume that the Lagrangian is of the form
\begin{equation}
\mathcal{L}=\mathcal{L}_{R}+\varepsilon\mathcal{L}_{R-breaking},
\label{LRexpl}
\end{equation}
where $\mathcal{L}_{R}$ preserves $R$-symmetry, the second term, $\mathcal{L}_{R-breaking}$,
is higher order in fields and breaks $R$-symmetry, and
$\varepsilon$ is parametrically small (we discuss why this should be shortly). Because $R$-symmetry
is broken explicitly by the second term, the Nelson-Seiberg theorem requires that
a global supersymmetry-preserving minimum must appear at order $1/\varepsilon$
away from the SUSY breaking one which now becomes metastable.
Note that this statement is completely general. Any attempt to mediate SUSY breaking
to gauginos even from models that initially
have no SUSY-preserving vacuum
results in the appearance of a global SUSY minimum.
Also the gaugino masses depend, as one would expect, on both the scale
of SUSY breaking \emph{and} the scale of $R$-symmetry breaking, whereas
the scalar masses depend only on the former.
(This point was used previously in \cite{split2}
in support of split SUSY \cite{split1}).
The gaugino masses are directly related
to $\varepsilon$ and hence to the stability of the metastable vacuum.

The second possibility is to break the tree-level $R$-symmetry spontaneously.
Spontaneous (rather than explicit) breaking of $R$-symmetry does not introduce
new global SUSY preserving minima. As such it does not destabilize the SUSY breaking
vacuum and does not require any fine-tuning of coefficients in the Lagrangian. At the same
time, gauginos do acquire masses.
This scenario, however, leads to a massless Goldstone mode of the spontaneously broken
$U(1)_{R}$ symmetry -- an \emph{$R$-axion problem}. In order to
avoid astrophysical (and experimental) bounds, the $R$-axion
should also acquire a mass. This means that
$R$-symmetry must also be explicitly broken and by the earlier arguments
this again means that the vacuum is metastable.
However in this case \cite{ADJK}
the gaugino mass is divorced from the size of \emph{explicit} $R$-breaking
$\varepsilon$ which now determines the $R$-axion mass instead. This
exhausts the logical possibilities and shows that, for a theory with a generic superpotential
where the Nelson-Seiberg theorem applies, massive gauginos
and massive $R$-axions imply metastability.

At this point the question arises as how to generate a Lagrangian
of the form \eqref{LRexpl}. Unless there is a compelling reason for the
smallness of $\varepsilon$, the Lagrangian
$\mathcal{L}_{R}$ is by definition non-generic,
and $\mathcal{L}_{R-breaking}$ may allow many couplings which are compatible
with the symmetries that one has to set to be small in order to avoid
too rapid decay of the metastable vacuum. One requires an \emph{almost} non-generic model,
broken by small operators, which in general seems unlikely. However, realistic and natural
gauge mediation models of this type were constructed in \cite{MN,AS}. The main idea of these models
is to break $R$-symmetry by operators which are suppressed by powers of $M_{Pl}.$
We will consider these models in Section {\bf 3}.

In \cite{ADJK} we suggested an alternative approach
where $\varepsilon$ is not induced by external $1/M_{Pl}$ corrections and where $R$-symmetry is broken spontaneously.
In the original ISS model \cite{ISS}, the Nelson-Seiberg theorem manifests itself in
a simple way: the theory has an exact $R$-symmetry at tree-level.
However the $R$-symmetry is anomalous and terms of the type $\varepsilon\mathcal{L}_{R-breaking}$
are generated dynamically  \cite{ISS}
without having to appeal to Planck suppressed
operators. Here $\varepsilon$ is a naturally small
parameter since it is generated non-perturbatively via instanton-like
configurations, which are naturally suppressed by the usual instanton factor
$e^{-8\pi^2/g^2} \ll 1.$ Hence, the non-genericity in these models is fully calculable and under control.
When, in addition to these non-perturbative effects, the
$R$-symmetry is also broken spontaneously by perturbative contributions, gauginos
receive sufficiently large masses
$m_{\rm gaugino} > 100$ GeV as required by their non-observation by current experiments.
At the same time the $R$-axion receives a mass from the anomalously induced
$R$-breaking terms.  (Note that a possible additional contribution to the
$R$-axion mass may arise when the theory is embedded in supergravity \cite{Bagger:1994hh}. However such noncalculable effects are suppressed.)

The spontaneous breaking of $R$-symmetry by radiative perturbative
corrections is easy to achieve \cite{DM,Shih:2007av}. For example, this happens \cite{ADJK}
when the basic ISS model is deformed by adding a baryon-like term to the superpotential.
This is the simplest deformation of the ISS model which preserves $R$-symmetry
at tree-level. At one-loop level this deformation causes the $R$-symmetry to break spontaneously,
while the $R$-axion gets a sufficiently large mass
$m_{\rm axion} > 100$ MeV to avoid astrophysical
constraints from the non-perturbative anomalous $R$-symmetry breaking \cite{ADJK}. No new global minima
appear other than those of the original ISS model, so the SUSY breaking
scale can be sufficiently low to be addressed at the LHC. These models
will be discussed in Section {\bf 2}.

{The paper is organised as follows. For convenience, in the following Section we recall the original ISS model \cite{ISS}.
In Section {\bf 2} we study gauge mediation with spontaneous $R$-symmetry breaking.}
Specifically, we concentrate on the \emph{direct} gauge mediation model \cite{ADJK} where the
Hidden plus Messenger sectors
consist of only the baryon-deformed ISS theory with $N_f=7$ flavours and
$N_c=5$ colours.
The resulting gaugino and sfermion soft masses are discussed in
Section {\bf 2.2}.
The mass term for the $R$-axion is generated by the nonperturbative ISS superpotential
\eqref{Wdyn}, as explained in \cite{ADJK}. In Section {\bf 2.3} we analyse the phenomenology of this class of models, which
turns out to be quite different from the usual gauge mediation scenarios \cite{GR}. The main reason for this difference
is the fact that $R$-symmetry is broken spontaneously by one-loop corrections, and as such the scale of $R$-breaking is naturally
smaller than the scale of SUSY-breaking, leading to the gaugino masses being
smaller than the scalar masses.
This is different from the usual gauge-mediation assumption that the
$R$-symmetry breaking is larger than the SUSY breaking. Thus generally, one expects a Hidden sector with spontaneous
 $R$-symmetry violation to interpolate between standard gauge mediation and split SUSY models \cite{split1,split2}.

In Section {\bf 3} we study the alternative scenario for metastable gauge mediation, formulated earlier
in Refs.~\cite{MN,AS}. These are models with an explicit messenger sector where the $R$-symmetry of the ISS sector here
is broken explicitly. As already mentioned, the reason why the effective $R$-symmetry breaking is weak in {this case is the fact that}
the messengers are coupled to the Hidden Sector fields only via $1/M_{Pl}$-suppressed
operators, cf. \eqref{WRff}.
In the limit where $M_{Pl} \to \infty$, both the $R$-symmetry and supersymmetry of the MSSM are exact,
since the ISS Hidden Sector decouples from the messengers.
As a result, in these models the effective $R$-symmetry breaking and the effective SUSY-breaking scales
in the Visible Sector are essentially the same. The generated gaugino and scalar soft mass terms are of the same order,
and the resulting phenomenology of models \cite{MN,AS} is largely of the usual form.

In both cases we will be treating the $\mus$ parameter of the MSSM as a free parameter.
As it is SUSY preserving it does not have to be determined by the ISS Hidden Sector, and we will
for this discussion have little to say about it: we will not address the question of why it
should be of order the weak scale, the so-called $\mu$ problem. However the corresponding
SUSY breaking parameter, $B$ or more precisely $B\mus$,
cannot consistently be taken to be a free parameter. It is determined by the
models at the messenger scale, and in both cases it is approximately zero, as will be explained
in detail.

\subsection{The ISS model -- summary}

In Ref.~\cite{ISS} Intriligator, Seiberg and Shih pointed out that metastable SUSY-breaking
vacua can arise naturally and dynamically in low-energy limits of supersymmetric gauge theories.
The simplest prototype model is SQCD with the gauge group $SU(N_c)$
and $N_f$ pairs of (anti)-fundamental quark supermultiplets $Q$, $\tilde{Q}$.
Metastable vacua $\vplus$ occur in this model when $N_f$ is in the `free magnetic range',
$N_c+1\le N_f \le \frac{3}{2} N_c.$ These vacua are apparent in the Seiberg dual formulation
of the theory, which has the advantage of being weakly coupled in the vicinity of $\vplus$.
The magnetic Seiberg dual of the ISS theory is given \cite{Seiberg1,Seiberg2} by the
$SU(N)_{mg}$ gauge theory, where $N=N_f-N_c,$ coupled to $N_f$ magnetic quark/anti-quark pairs
$\varphi$, $\tilde{\varphi}$. The tree-level superpotential of the magnetic theory is of the
form,
\begin{equation}
W_{\rm cl} =\Phi_{ij}\,\varphi_{i}\cdot\tilde{\varphi_{j}}-\mu_{ij}^{2}\Phi_{ji}
\label{Wcl}
\end{equation}
where $i,j=1...N_f$ are flavour indices and $\Phi_{ij}$ is the gauge-singlet meson superfield,
which is related to the original electric quarks via $\Phi_{ij} \propto \Lambda^{-1} Q_i \cdot\tilde{Q}_j$
and $\Lambda$ is the dynamical scale of the ISS theory \cite{ISS}.
The matrix $\mu_{ij}^2$ (which can be diagonalised without loss of generality) arises from
the masses of electric quarks, $\mu^2_{ii}= \Lambda m_{Q_i},$ and is taken to be much smaller than
the UV cutoff of the magnetic theory, $\mu \ll \Lambda.$ This magnetic theory is free and calculable in the
IR and becomes strongly coupled in the UV where one should use instead the electric Seiberg dual, i.e.
the original $SU(N_c)$ SQCD which is asymptotically free.

The usual holomorphicity arguments imply that the superpotential
\eqref{Wcl} receives no corrections in perturbation theory. However, there is a non-perturbative contribution
to the full superpotential of the theory, $W=W_{\rm cl} + W_{\rm dyn},$ which
is generated dynamically  \cite{ISS} and is given by
\be
W_{\rm dyn}\, =\, N\left(  \frac{\det_{\sst \Nf} \Phi}{\Lambda^{\Nf-3N}}\right)^\frac{1}{N}
\label{Wdyn}
\ee
The authors of \cite{ISS} have studied the vacuum structure of the theory and established the
existence of the metastable vacuum $\vplus$ with non-vanishing vacuum energy $V_{+}$
as well as the (set of $N_c$) SUSY preserving stable vacua $\vzero$.

This supersymmetry breaking vacuum $\vplus$ originates from the so-called
rank condition, which implies that there are no solutions to the F-flatness
equation\footnote{Equation \eqref{rank-cond}
can only be satisfied for a rank-$N$ submatrix of the $N_f \times N_f$ matrix
$F_{\Phi}$.}
\be
F_{\Phi_{ij}}\, =\, (\varphi_{i}\cdot\tilde{\varphi}_{j}-\mu_{ij}^2)\, =\, 0
\label{rank-cond}
\ee
for the classical superpotential $W_{\rm cl}.$
The SUSY preserving vacuua
\eqref{vac0} appear by allowing the meson $\Phi$ to develop a VEV
which is stabilised by the non-perturbative superpotential \eqref{Wdyn} and
can be interpreted in the ISS model as a non-perturbative
or dynamical restoration of supersymmetry \cite{ISS}.
The lowest lying SUSY-breaking
vacuum $\vplus$ is characterised by
\be
\langle \p \rangle =\, \langle \pt^T \rangle = \, \left(\begin{array}{c}
{\rm diag}(\mu_1,\ldots,\mu_N) \\ 0_{\Nf-N}\end{array}\right) \ , \quad
\langle \Phi \rangle = \, 0 \ ,
\qquad V_+ = \sum_{i=N+1}^{N_f}|\mu_i^4|.
\label{vac+}
\ee
Here $\mu_i$ are the ordered eigenvalues $\mu$ matrix, such that
$|\mu_1| \ge |\mu_2| \ge \ldots \ge |\mu_{N_f}|.$ In this way, the vacuum energy $V_+$
above receives contributions from $(\Nf-N)$ of the smallest $\mu$'s while the VEV $\langle \p \rangle$
is determined by the largest $\mu$'s.

The SUSY-preserving vacuum
$\vzero$ is described by\footnote{In fact there are precisely $\Nf-N=\Nc$ of such vacua
differing by the phase $e^{2\pi i/(\Nf-N)}$ as required by the Witten index of the electric ISS theory.}
\be
\langle \p \rangle =\, \langle \pt^T \rangle = \, 0 \ , \quad
\langle \Phi \rangle = \, \left(\frac{\Lambda}{\mu}\right)^{\frac{N_f-N}{N_f-3N}} \mu \, \uno_{\Nf} \ ,
\qquad \qquad V_0 = \, 0,
\label{vac0}
\ee
where for simplicity we have specialised to the degenerate case, $\mu_{ij}=\mu \delta_{ij}.$
For $\mu/\Lambda \ll 1$ the metastable vacuum is exponentially long-lived and the lifetime of
$\vplus$ can easily be made much longer than the age of the Universe.
One very attractive feature of these models is
that at high temperatures the SUSY breaking
vacua are dynamically favoured over the SUSY preserving ones \cite{ACJK}.
This is
because the metastable ISS-type vacua have more light degrees of freedom, so the early Universe would
naturally have been driven into them \cite{ACJK,heat2,heat3,heat4}.

Other recent investigations of metastable SUSY-breaking applied to model building include
Refs.~\cite{Argurio:2006ny,Kitano:2006xg,Csaki:2006wi,Abel:2007uq,Abel:2007zm,Ferretti:2007ec,Brummer:2007ns,Essig:2007xk,Haba:2007rj,Cheung:2007es,Heckman:2007zp}

\section{Gauge mediation with Spontaneous $R$-symmetry breaking}

The metastable model building paradigm makes it relatively
easy to construct models with dynamically broken supersymmetry. The simplicity
of the resulting models now compels us to consider the attractive possibility
of \emph{direct} gauge mediation, whereby matter fields of the SUSY-breaking sector carry charges under
the gauge groups of the Standard Model and there is no need for a separate messenger sector.
In ordinary gauge mediation, the details
of SUSY breaking are generally `hidden' from the matter sector, with the most
important phenomenological features arising from the messenger particle content.
The elegance of direct gauge mediation models lies in their compactness and
predictivity. Previously direct mediation of metastable SUSY breaking was considered
in this context
in Refs.~\cite{Kitano:2006xg,Csaki:2006wi} and \cite{ADJK}.

The essential difference between the direct gauge mediation of SUSY-breaking and the models with explicit
messengers \cite{MN,AS} is that the `direct messengers' form an integral part of the Hidden
ISS sector, and as such, their interactions with the SUSY-breaking VEVs are not suppressed by inverse powers
of $M_{Pl}$. This means that the $R$-symmetry of the SUSY-breaking sector (required by the existence of the
SUSY-breaking vacuum) cannot be an accidental symmetry which is violated in the full theory only by
$1/M_{Pl}$
corrections, as in \cite{MN,AS}.
On the other hand, any large explicit violations of $R$-symmetry in
the
full theory will necessarily destabilise
the SUSY-breaking metastable vacuum. Thus, it was proposed in our earlier paper \cite{ADJK}
that the $R$-symmetry must be \emph{spontaneously} broken by radiative corrections arising from the
Coleman-Weinberg potential. In this case the Nelson-Seiberg theorem does not force upon us
a nearby supersymmetric vacuum and at the same time non-zero gaugino masses can be generated since the
$R$-symmetry is broken.

We will show below that in this approach the direct gauge mediation scenarios give
phenomenology quite distinct from the usual gauge mediation scenarios \cite{GR}.

\subsection{The baryon-deformed ISS model \cite{ADJK}}

To break supersymmetry we take an ISS model with $N_{c}=5$ colours and
$N_{f}=7$ flavours, which has a magnetic dual description as an $SU(2)$ theory,
also with $N_{f}=7$ flavours.\footnote{These are the minimal allowed values of $N_c$ and
$N_f$ which lead to a non-trivial -- in this case $SU(2)$ -- magnetic gauge group.}
 Following \cite{ADJK} we now deform this theory by the
addition of a baryonic operator.
The superpotential of the theory is given by
\begin{equation}
W=\Phi_{ij}\varphi_{i}.\tilde{\varphi_{j}}-\mu_{ij}^{2}\Phi_{ji}+m\varepsilon_{
ab}\varepsilon_{rs}\varphi_{r}^{a}\varphi_{s}^{b}\,\,
\label{Wbardef}
\end{equation}
where $i,j=1...7$ are flavour indices, $r,s=1,2$ run over the first
two flavours only, and $a,b$ are $SU(2)$ indices.
This is the superpotential of ISS with the
exception of the last term which is a baryon of the magnetic $SU(2)$ gauge group. Note that
the 1,2 flavour indices and the 3...7 indices have a different status
and the full flavour symmetry $SU(7)_f$ is broken explicitly to $SU(2)_{f}\times SU(5)_{f}$.
The $SU(5)_{f}$ factor is gauged separately and identified
with the parent $SU(5)$ gauge group of the standard model.
The matter field decomposition under the magnetic $SU(2)_{gauge} \times SU(5)_{f}\times SU(2)_{f}$ and their $U(1)_R$ charges are given
in Table~\ref{fieldstable}.
\begin{table}
\begin{center}
\begin{tabular}{|c|c|c|c|c|}
\hline
&{\small $SU(2)$} &
{\small $SU(2)_f$}&
$SU(5)_f$&
{\small $U(1)_{R}$}\tabularnewline
\hline
\hline
$\Phi_{ij}\equiv\left(\begin{array}{cc}
Y & Z\\
\tilde{Z} & X\end{array}\right)$&
{\bf 1}&
$\left(\begin{array}{cc}
Adj +{\bf 1} & \bar\square\\
\square & {\bf 1}\end{array}\right)$&
$\left(\begin{array}{cc}
{\bf 1} & \square\\
\bar{\square} & Adj+{\bf 1}\end{array}\right)$&
2
\tabularnewline
\hline
{\small $\varphi\equiv\left(\begin{array}{c}
\phi\\
\rho\end{array}\right)$}&
$\square$&
$\left(\begin{array}{c}
\bar{\square}\\ {\bf 1} \end{array}\right)$&
$\left(\begin{array}{c}
{\bf 1}\\
\bar{\square}\end{array}\right)$&
$1$
\tabularnewline
\hline
{\small $\tilde{\varphi}\equiv\left(\begin{array}{c}
\tilde{\phi}\\
\tilde{\rho}\end{array}\right)$}&
$\bar\square$&
$\left(\begin{array}{c}
\square\\ {\bf 1} \end{array}\right)$&
$\left(\begin{array}{c}
{\bf 1}\\
\square\end{array}\right)$&
$-1$\tabularnewline
\hline
\end{tabular}
\end{center}
\caption{We list matter fields and their decomposition under the gauge $SU(2)$, the flavour $SU(2)_f \times SU(5)_f$ symmetry,
and their charges under the $R$-symmetry
of the model in \eqref{Wbardef}.
\label{fieldstable}}
\end{table}

It is known that the $R$-symmetry of the ISS SQCD manifests itself only as an approximate symmetry
of the magnetic formulation which is broken explicitly in the electric theory by the mass terms of electric quarks $m_Q$.
(It is also broken anomalously, but this is already accounted for by the dynamical superpotential \eqref{Wdyn}.)
In the Appendix we point out that the $R$-symmetry is broken in the electric theory in a controlled way
by small parameter, $m_Q /\Lambda = \mu^2 /\Lambda^2 \ll 1$. As such the $R$-symmetry is preserved to that order in the
superpotential.

Thanks to the baryon deformation, the model
has $R$-charges that are not 0 or 2. As discussed in Ref.~\cite{Shih:2007av}
this condition is necessary for Wess-Zumino models
to spontaneously break $R$-symmetry. Therefore, our model allows for spontaneous $R$ symmetry breaking
and we have shown in \cite{ADJK} that this does indeed happen.
We also stress that our baryon deformation is the leading order deformation of
the ISS model that is allowed by $R$-symmetry of the full theory imposed at the Lagrangian level.
As explained in the Appendix, this is a self-consistent approach since $R$-symmetry breaking
in the electric theory is controlled by a small parameter.
Terms quadratic in the
meson $\Phi$ that could arise from lower dimensional irrelevant operators
in the electric theory are forbidden by $R$-symmetry.
Thus, our deformation is described by a \emph{generic}
superpotential and \eqref{Wbardef} gives its leading-order terms.

Using the $SU(2)_{f}\times SU(5)_{f}$ symmetry, the matrix
$\mu_{ij}^{2}$ can be brought to a diagonal form
\begin{equation}
\mu_{ij}^{2}=\left(\begin{array}{cc}
\mu^{2}\mathbf{I}_{2} & 0\\
0 & \hat{\mu}^{2}\mathbf{I}_{5}\end{array}\right).
\end{equation}
We will assume that $\mu^{2}> \hat{\mu}^{2}$. The parameters $\mu^{2}$,
$\hat{\mu}^{2}$ and $m$ have an interpretation in terms of the electric
theory: $\mu^{2}\sim\Lambda m_{Q}$ and $\hat{\mu}^{2}\sim\Lambda\hat{m}_{Q}$
come from the electric quark masses $m_{Q}$, $\hat{m}_{Q}$, where
$\Lambda$ is the ISS scale. The baryon operator can
be identified with a corresponding operator in the electric theory.
Indeed the mapping from baryons $B_{E}$ in the electric theory to
baryons $B_{M}$ of the magnetic theory, is $B_{M}\Lambda^{-N}\leftrightarrow
B_{E}\Lambda^{-N_{c}}$
(we neglect factors of order one). Thus one expects
\begin{equation}
m\sim
M_{Pl}\left(\frac{\Lambda}{M_{Pl}}\right)^{2N_{c}-N_{f}}=\frac{\Lambda^{3}}{M_{Pl}^
{2}},
\label{mbardef}
\end{equation}
where $M_{Pl}$ represents the scale of new physics in the electric
theory at which the irrelevant operator $B_{M}$ is generated.

As explained in \cite{ADJK},
this theory has a classical runaway direction $\langle \tilde\varphi \rangle \to \infty$
(with $\langle \tilde\varphi \rangle \langle \varphi \rangle$ fixed)
to a non-supersymmetric vacuum.
The quantum dynamics, namely the one-loop Coleman-Weinberg potential \cite{Coleman:1973jx},
\be
V_{\mathrm{eff}}^{(1)}\!=\!\frac{1}{64\pi^2}\,\mathrm{STr}\,\M^4\log\frac{\M^2}{
\Lambda^2_{UV}}\,
\!\equiv\frac{1}{64\pi^2}\left( \Tr\, m_{sc}^4\log\frac{m_{sc}^2}{\Lambda^2_{UV}}-2\,\Tr\,
  m_{f}^4\log\frac{m_{f}^2}{\Lambda^2_{UV}} +3\, \Tr\,
m_v^4\log\frac{m_v^2}{\Lambda^2_{UV}}\right)\label{CW}
\ee
stabilises the runaway
at a point which breaks both supersymmetry and
$R$-symmetry, thus creating a meta-stable vacuum state. ($m_{sc}$, $m_f$ and $m_v$ on the RHS denote mass matrices of all
relevant scalar, fermion and vector fields in the model and $\Lambda_{UV}$ is traded for a renormalisation scale
at which the couplings are defined.)
We parameterise the classically pseudo-Goldstone and runaway VEVs by
\bea
\label{phivevs}
\vev{\tilde\phi} &=& \xi\,\mathbf{I}_{2}\quad\quad\quad\quad\,\,\vev{\phi}=\kappa\,\mathbf{I}_{2}\\
\label{yvevs}
\vev{Y} &=& \eta\,\mathbf{I}_{2}\quad\quad\quad\quad\vev{X}=\chi\,\mathbf{I}_{5}.
\eea
These are the most general vevs consistent with the tree level minimization.
It can be checked that at one loop order all other field vevs are zero in the lowest perturbative vacuum.
By computing the masses of all fluctuations about this valley we can go about
constructing the one-loop effective potential from eq. \eqref{CW}.
We have done this numerically using {\em{Vscape}} program of Ref.~\cite{vandenBroek:2007kj}.
Table~\ref{vscapetable} gives a sample points showing the VEVs stabilised by the one
loop effective potential.

\begin{table}[h] \begin{center}
\begin{tabular}{|c|c|c|c|c|c|} \hline $\mu$ & $m$ & $\xi$ & $\kappa$ & $\eta$ & $\chi$ \\
\hline \hline 10 & 0.3 & 41.0523 & 2.43592 & $-0.035477$ & $-1.761261$ \\
\hline 1.1 & 0.3 & 2.1370 & 0.566214 & $-0.148546$ & $-0.083296$ \\
\hline 1.01 & 0.3 & 1.8995 & 0.537043 & $-0.155796$ & $-0.073474$ \\
\hline 1.003 & 0.3 & 1.8809 & 0.534848 & $-0.157752$ & $-0.072738$ \\
\hline \hline \end{tabular} \end{center} \caption{ Stabilised VEVs from Vscape for various parameter points.
All values are given in units of $\hat{\mu}$. \label{vscapetable}} \end{table}

To summarise, we have identified a SUSY-breaking vacuum of the deformed ISS model, which also breaks $R$-symmetry
spontaneously via radiative corrections.
This is a long-lived metastable vacuum. The SUSY-preserving vacua of this model are only those generated by the
non-perturbative suprepotential,
\begin{equation}
W_{np}=2\Lambda^{3}\left[\det\left(\frac{\Phi}{\Lambda}\right)\right]^{\frac{1}{2}}.
\end{equation}
Adapting the supersymmetric vacuum solution from the ISS model to our case with $\mu>\hat{\mu}$ we find,
\begin{equation}
\varphi=0,\quad\tilde{\varphi}=0,\quad\eta=\hat{\mu}^{2}\mu^{-\frac{6}{5}}
\Lambda^{\frac{1}{5}}\quad \chi=\mu^{\frac{4}{5}}\Lambda^{\frac{1}{5}}.
\end{equation}
Note that the supersymmetric minimum lies at $\varphi=\tilde{\varphi}=0$ and is completely
unaffected by the baryon
deformation. As we are not breaking $R$-symmetry explicitly, no other supersymmetric vacua are generated,
and, as a result, the decay rate of our metastable vacuum is exponentially small as in the original ISS model.

\subsection{Direct gauge mediation and generation of gaugino and sfermion masses}

As mentioned earlier, the $SU(5)_{f}$ symmetry of the superpotential \eqref{Wbardef} is gauged and identified
with the parent $SU(5)$ of the MSSM sector. This induces direct gauge mediation of SUSY breaking from the metastable
vacuum of the Hidden ISS sector to the MSSM. The Hidden sector matter fields $\rho$, $\tilde{\rho}$, $Z$, $\tilde{Z}$ and $X$
are charged under the $SU(5)$ and serve as direct messengers. These induce all the soft SUSY-breaking terms in the MSSM
sector, including gaugino and sfermion masses.

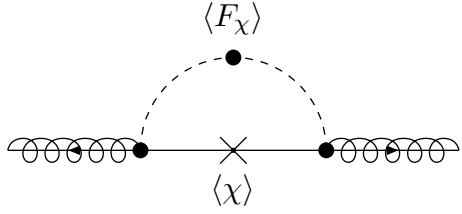
\begin{figure}[t]
\begin{center}
\begin{picture}(200,120)
\SetOffset(0,20)
\Gluon(0,0)(50,0){4.5}{5}
\ArrowLine(50,0)(0,0)
\Line(50,0)(120,0)
\Gluon(120,0)(170,0){4.5}{5}
\Vertex(50,0){3}
\Vertex(120,0){3}
\DashCArc(85,0)(35,0,180){3}
\ArrowLine(120,0)(170,0)
\Vertex(85,35){3}
\Text(85,50)[c]{\scalebox{1.1}[1.1]{$\langle F_{\chi}\rangle$}}
\Vertex(85,0){1}
\Line(80,5)(90,-5)
\Line(90,5)(80,-5)
\Text(85,-15)[c]{\scalebox{1.1}[1.1]{$\langle \chi\rangle$}}
\end{picture}
\end{center}
\vspace*{-0.5cm} \caption{\small One-loop contribution to the
gaugino masses. The dashed (solid) line is a bosonic (fermionic) messenger.
The blob on the scalar line indicates an insertion of $\langle F_{\chi}\rangle$ into the propagator
of the scalar messengers and the cross denotes an
insertion of the $R$-symmetry breaking VEV into the propagator of the fermionic messengers.}\label{gauginofig}
\end{figure}

Gaugino masses are generated at one loop order (cf. Fig. \ref{gauginofig}). The fields propagating in the loop are
fermion and scalar components of the direct mediation `messengers' $\rho$, $\tilde{\rho}$
and $Z$, $\tilde{Z}$.\footnote{The adjoint part in
$X$ is also charged under the standard model gauge groups and therefore, in principle, can also mediate SUSY-breaking.
However, at tree-level $X$ does not couple to the supersymmetry breaking $F$-term,
and its fermionic and bosonic components have identical (zero) mass.
This degeneracy is only lifted at the one-loop level by the Coleman-Weinberg potential. We therefore neglect the contribution
from $X$ which we expect to be subdominant.}
Since
gaugino
masses are forbidden by $R$-symmetry one crucial ingredient in their generation is the presence of
a non-vanishing $R$-symmetry breaking VEV - in our case $\langle\chi\rangle$ generated by the non-vanishing baryon deformation $m$.

In contrast to the gaugino masses $m_{\lambda}$, sfermion masses $m_{\tilde{f}}$ are not protected by $R$-symmetry. Hence, as
long as supersymmetry remains broken, we can have non-vanishing
sfermion masses even in
the
absence of an $R$-symmetry breaking VEV. In our model that means that the sfermion masses are non-vanishing even in the case of
a vanishing baryon deformation.
This shows that in a general (gauge) mediation scenario sfermion and gaugino masses are generated by quite different mechanisms.
Accordingly the simple relation $m_{\lambda}\sim m_{\tilde{f}}$ does not necessarily hold in general gauge mediation scenarios. Indeed our model
is an explicit example where it fails.

Let us now turn to the practical evaluation of the gaugino masses.
For fermion components of the messengers,
\begin{equation}
\psi =  (\rho_{ia}\,,Z_{ir})_{ferm} , \quad
\tilde{\psi}  =  (\tilde{\rho}_{ia}\,,\tilde{Z}_{ir})_{ferm},
\end{equation}
the mass matrix is given by
\begin{equation}
m_{f}=\mathbf{I}_{5}\otimes\mathbf{I}_{2}\otimes\left(\begin{array}{cc}
\chi & \xi\\
\kappa & 0\end{array}\right).
\label{mferms}
\end{equation}
We can also assemble the relevant scalars into
\begin{equation}
\label{Sdef}
S=(\rho_{ia},Z_{ir},\tilde{\rho}_{ia}^{*},\tilde{Z}_{ir}^{*})_{sc},
\end{equation}
and for the corresponding scalar mass-squared matrix we have
\begin{equation} \label{msc}
m_{sc}^{2}=\mathbf{I}_{5}\otimes\mathbf{I}_{2}\otimes\left(\begin{array}{cccc}
|\xi|^{2}+|\chi|^{2} & \chi^{*}\kappa & -\hat{\mu}^{2} &
\eta\,\kappa \\
\chi\,\kappa^{*} & |\kappa|^{2} &
\,\xi\,\eta+2m\kappa\, & 0\\
-\hat{\mu}^{2} & (\xi\eta)^{*}+2m\kappa^{*}  &
|\kappa|^{2}+|\chi|^{2} & \chi\,\xi^{*} \\
\eta{}^{*}\kappa^{*} & 0 & \chi^{*}\xi & |\xi|^{2}\end{array}\right).
\end{equation}

Gaugino masses arise from the one-loop diagram in Fig. \ref{gauginofig}.
To evaluate the diagram it is convenient to diagonalize the non-diagonal mass
terms \eqref{mferms}, \eqref{msc} using unitary matrices,
\begin{eqnarray}
\hat{m}_{sc}^{2} & = & Q^{\dagger}m_{sc}^{2}Q\\
\hat{m}_{f} & = & U^{\dagger}m_{f}V.
\end{eqnarray}
The fields in the new basis are given by,
\begin{eqnarray}
\hat{S} & = & S.Q\\
\hat{\psi}_{+} & = & \psi.U\\
\hat{\psi}_{-} & = & \tilde{\psi}.V^{*}.
\end{eqnarray}

In order to calculate the gaugino mass, we need the gauge interaction
terms given by
\begin{eqnarray}
\label{gaugeint}
\mathcal{L} & \supset & i\sqrt{2}g_{A}\lambda_{A}(\psi_{1}T^{A}S_{1}^{*}+\psi_{2}T^{A}S_{2}^{*}
+\tilde{\psi}_{1}T^{*A}S_{3}+\tilde{\psi}_{2}T^{*A}S_{4})+H.C.\\
 & = & i\sqrt{2}g_{A}\lambda_{A}(\hat{\psi}_{+i}\hat{S}_{k}^{*}(U_{i1}^{\dagger}Q_{1k}+U_{i2}^{\dagger}Q_{2k})+
 \hat{\psi}_{-i}\hat{S}_{k}(Q_{k3}^{\dagger}V_{1i}+Q_{k4}^{\dagger}V_{2i}))+H.C,
 \label{gaugeint2}
\end{eqnarray}
where we have expressed everything in terms of the
mass eigenstates in the second line.

Using the gauge interactions Eq. \eqref{gaugeint2}, the diagram in Fig. \ref{gauginofig} contributes to gaugino masses
as follows\footnote{More precisely, in evaluating \eqref{gauginomass}, we use the diagram in Fig. \ref{gauginofig}
without explicit insertions of $\langle F_{\chi}\rangle$ and $\langle {\chi}\rangle$ in the messenger propagators.
In the loop we use mass-eigenstate propagators and insert the
diagonalisation matrices at the vertices.
Appropriate dependence on $\langle F_{\chi}\rangle$ and $\langle {\chi}\rangle$ is
automatically introduced by the diagonalisation matrices.},
\begin{equation}
m_{\lambda_A}=2g_{A}^{2}\, Tr(T^{A}T^{B})\sum_{ik}(U_{i1}^{\dagger}Q_{1k}+U_{i2}^{\dagger}Q_{2k})(Q_{k3}^{\dagger}V_{1i}+Q_{k4}^{\dagger}V_{2i})
\, I(\hat{m}_{f,i},\hat{m}_{sc,k})
\label{gauginomass}
\end{equation}
where the 1-loop integral $I$ evaluates to
\begin{equation}
I(a,b)=\frac{-a(\eta+1)}{16\pi^2}+\frac{1}{16\pi^2}\frac{a}{a^2-b^2}(a^2\log(a^2/\Lambda^2)-b^2\log(b^2/\Lambda^2)),
\end{equation}
\begin{equation}
\eta=\frac{2}{4-D}+\log(4\pi)-\gamma_{E}.
\end{equation}
$I(a,b)$ is UV-divergent, but the divergences cancel in the sum over eigenstates as
they should.

\begin{figure}[t]
\begin{center}
\subfigure[]{\begin{picture}(210,100)
\Text(-22,110)[l]{\scalebox{1.6}[1.6]{$\frac{m_{1/2}}{\hat{\mu}}$}}
\Text(180,-5)[l]{\scalebox{1.3}[1.3]{$m/\hat{\mu}$}}
\includegraphics[width=.43\textwidth]{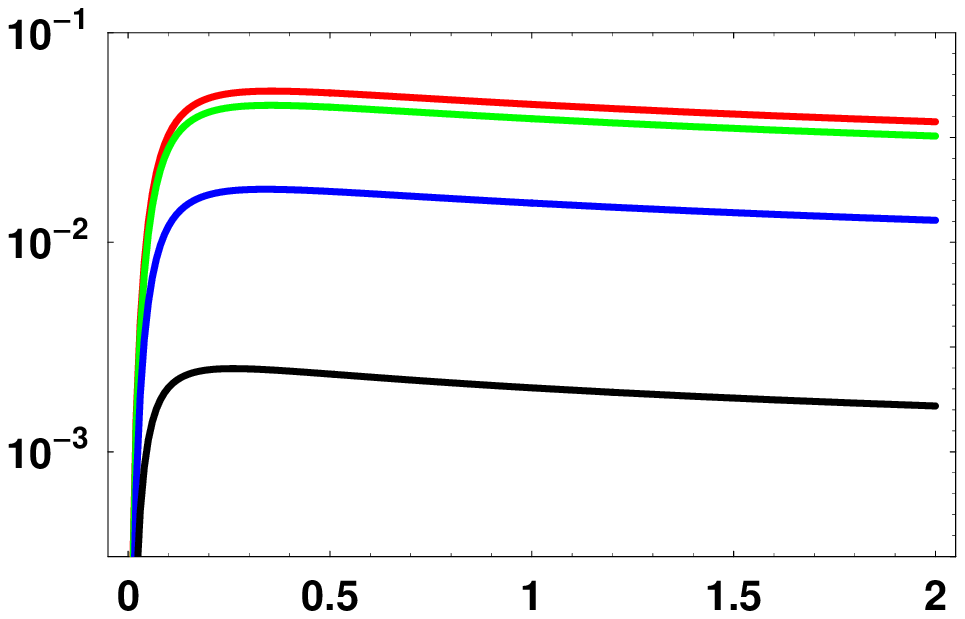}
\end{picture}\label{gauginomassnum}}
\hspace{1.5cm}
\subfigure[]{\begin{picture}(210,100)
\Text(-8,110)[l]{\scalebox{1.6}[1.6]{$\frac{m_{0}}{\hat{\mu}}$}}
\Text(190,-5)[l]{\scalebox{1.3}[1.3]{$m/\hat{\mu}$}}
\includegraphics[width=.43\textwidth]{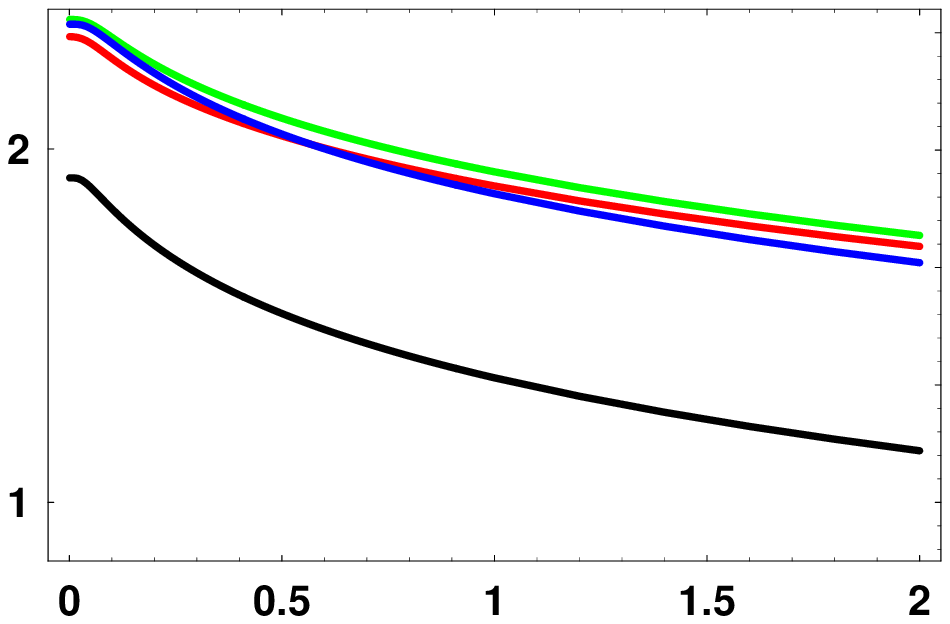}
\end{picture}\label{sfermionmassnum}}
\end{center}
\caption{\small Gaugino mass scale, $m_{1/2}$, and sfermion mass scale, $m_{0}$,  as functions of the baryon deformation $m$,
for various values of $\mu$: red ($\mu=1.003$), green ($\mu=1.01$), blue ($\mu=1.1$) and black ($\mu=1.5$).
The mass scales $m_{1/2}$ and $m_{0}$ are defined in Eqs.~\eqref{mhalfdef},\eqref{mzerodef}.}
\label{gauginomassnumresult}
\end{figure}

Keeping the SUSY-breaking scale $\hat\mu$ fixed we can now study the dependence of the gaugino mass
on the two remaining parameters $\mu$ and $m$.
The VEVs $\xi$, $\kappa$, $\eta$ and $\chi$ are generated from minimizing the effective potential,
as above.
The results are shown in Fig.~\ref{gauginomassnumresult}(a).

For completeness we note that
the usual analytic expression for the gaugino mass valid to the
leading order ${\cal O}(F_{\chi})$, is of little use for us.
It comes from magnetic quarks,
$\rho$ and $\tilde{\rho}$, propagating in the loop, as shown in Fig.
\ref{gauginofig}, and takes form  $m^{(1)}_{\lambda_{A}} \sim \frac{g_{A}^{2}}{16\pi^{2}}\, F_{\chi}\,
(m_{f})^{-1}_{11}.$ However it trivially
vanishes in our model, and one needs to go to order $F_{\chi}^3$ to find
a non-vanishing contribution. This effect was first pointed out in Ref.~\cite{Izawa:1997gs}:
the zero
element in the lower right corner of the fermion mass matrix
\eqref{mferms} implies that $(m_{f})^{-1}_{11} =0$ and hence $m^{(1)}_{\lambda_{A}}=0$.

\begin{figure}[t]
\begin{center}
\subfigure[]{
\scalebox{0.6}[0.6]{\begin{picture}(174,120)
\SetOffset(0,20)
\DashLine(0,0)(50,0){1}
\DashLine(50,0)(120,0){1}
\DashLine(120,0)(170,0){1}
\Vertex(50,0){3}
\Vertex(120,0){3}
\Vertex(85,40){3}
\Gluon(50,0)(85,40){4}{6}
\Gluon(85,40)(120,0){4}{6}
\DashCArc(85,60)(20,0,360){3}
\end{picture}}\label{twoloop1}}
\subfigure[]{
\scalebox{0.6}[0.6]{\begin{picture}(174,120)
\SetOffset(0,20)
\DashLine(0,0)(50,0){1}
\DashLine(50,0)(120,0){1}
\DashLine(120,0)(170,0){1}
\Vertex(50,0){3}
\Vertex(120,0){3}
\Gluon(50,0)(60,40){4}{5}
\Gluon(110,40)(120,0){4}{5}
\Vertex(60,40){3}
\Vertex(110,40){3}
\DashCArc(85,30)(28,30,150){3}
\DashCArc(85,50)(28,210,330){3}
\end{picture}}\label{twoloop2}}
\subfigure[]{
\scalebox{0.6}[0.6]{\begin{picture}(174,120)
\SetOffset(0,20)
\DashLine(0,0)(50,0){1}
\DashLine(50,0)(120,0){1}
\DashLine(120,0)(170,0){1}
\Vertex(85,0){3}
\Vertex(85,40){3}
\GlueArc(85,20)(20,90,270){-4}{6.5}
\GlueArc(85,20)(20,270,450){-4}{6.5}
\DashCArc(85,60)(20,0,360){3}
\end{picture}}\label{twoloop3}}
\subfigure[]{
\scalebox{0.6}[0.6]{\begin{picture}(174,120)
\SetOffset(0,20)
\DashLine(0,0)(50,0){1}
\DashLine(50,0)(120,0){1}
\DashLine(120,0)(170,0){1}
\Vertex(85,0){3}
\GlueArc(92,38)(35,175,260){4}{5}
\GlueArc(78,38)(35,280,365){4}{5}
\Vertex(60,40){3}
\Vertex(110,40){3}
\DashCArc(85,30)(28,30,150){3}
\DashCArc(85,50)(28,210,330){3}
\end{picture}}\label{twoloop4}}
\subfigure[]{
\scalebox{0.6}[0.6]{\begin{picture}(174,120)
\SetOffset(0,20)
\DashLine(0,0)(50,0){1}
\DashLine(50,0)(120,0){1}
\DashLine(120,0)(170,0){1}
\Vertex(85,0){3}
\GlueArc(92,38)(35,175,260){4}{5}
\GlueArc(78,38)(35,280,365){4}{5}
\Vertex(60,40){3}
\Vertex(110,40){3}
\CArc(85,30)(28,27,153)
\CArc(85,50)(28,207,333)
\end{picture}}\label{twoloop5}}
\subfigure[]{
\scalebox{0.6}[0.6]{\begin{picture}(174,120)
\SetOffset(0,20)
\DashLine(0,0)(50,0){1}
\DashLine(50,0)(120,0){1}
\DashLine(120,0)(170,0){1}
\Vertex(50,0){3}
\Vertex(120,0){3}
\Gluon(50,0)(60,40){4}{5}
\Gluon(110,40)(120,0){4}{5}
\Vertex(60,40){3}
\Vertex(110,40){3}
\DashCArc(85,30)(28,30,150){3}
\DashCArc(85,50)(28,210,330){3}
\end{picture}}\label{twoloop6}}
\subfigure[]{
\scalebox{0.6}[0.6]{\begin{picture}(174,120)
\SetOffset(0,20)
\DashLine(0,0)(50,0){1}
\DashLine(50,0)(120,0){1}
\DashLine(120,0)(170,0){1}
\Vertex(50,0){3}
\Vertex(120,0){3}
\DashCArc(85,-17)(40,30,150){3}
\DashCArc(85,17)(40,-30,-150){3}
\end{picture}}\label{twoloop7}}
\subfigure[]{
\scalebox{0.6}[0.6]{\begin{picture}(174,120)
\SetOffset(0,20)
\DashLine(0,0)(50,0){1}
\DashLine(50,0)(120,0){1}
\DashLine(120,0)(170,0){1}
\Vertex(50,0){3}
\Vertex(120,0){3}
\Gluon(50,0)(50,40){4}{5}
\ArrowLine(50,0)(50,40)
\Gluon(120,40)(120,0){4}{5}
\ArrowLine(120,40)(120,0)
\Vertex(50,40){3}
\Vertex(120,40){3}
\DashCArc(85,40)(35,0,180){3}
\Line(50,40)(120,40)
\end{picture}}\label{twoloop8}}
\end{center}
\vspace*{-0.5cm} \caption{\small Two-loop diagrams contributing to the sfermion masses.
The long dashed (solid) line is a bosonic (fermionic) messenger.
Standard model sfermions are depicted by short dashed lines.}\label{sfermionfig}
\end{figure}
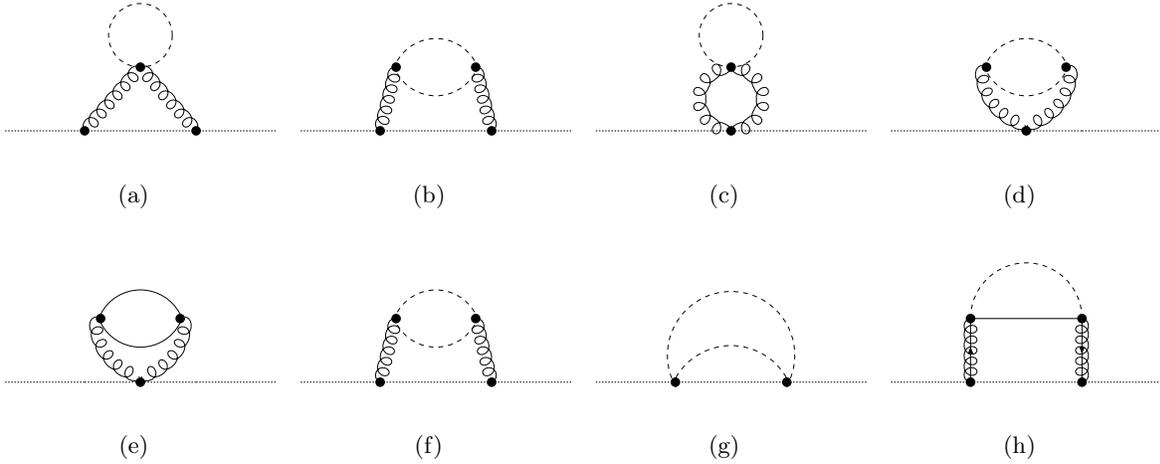

Having determined
the
gaugino masses in Eq.~\eqref{gauginomass} and Fig.~\ref{gauginomassnumresult}(a),
we now turn to the generation of the masses for the sfermions of the supersymmetric standard model.
Here we will closely follow the calculation in Ref.~\cite{Martin:1996zb} adapted to our more general set of messenger particles.
As already mentioned at the beginning of this section, sfermion masses are generated by a different mechanism than the scalar
masses. Indeed they are generated by
the
two-loop diagrams shown in Fig. \ref{sfermionfig}.
In \cite{Martin:1996zb} the contribution of these diagrams to the sfermion masses
was
determined to be,
\begin{equation}
m^{2}_{\tilde{f}}=\sum_{mess.} \sum_{a} g^{4}_{a} C_{a} S_{a}(mess.)[{\rm{sum}}\,\,{\rm{of}}\,\,{\rm{graphs}}],
\label{sfermionmass}
\end{equation}
where we sum over all gauge groups under which the sfermion is charged, $g_{a}$ is the corresponding gauge coupling,
$C_{a}=(N^{2}_{a}-1)/(2N_{a})$ is the quadratic Casimir and $S_{a}(mess.)$ is the Dynkin index of the messenger fields
(normalized to $1/2$ for fundamentals).

In the following we will only describe the new features specific to the messenger fields of our direct mediation model.
The explicit expressions for the loop integrals and
the algebraic prefactors resulting from the $\gamma$-matrix algebra etc. can be found in the appendix of \cite{Martin:1996zb}.
To simplify the calculation we also neglect the masses of the MSSM fields relative to the messenger masses.

As in the calculation of the gaugino mass we use the propagators in the diagonal form and insert the
diagonalisation matrices directly at the vertices.
For the diagrams \ref{twoloop1} to \ref{twoloop6} we have closed loops of purely bosonic or purely fermionic
mass eigenstates of our messenger fields.
It is straightforward to check that in this case the unitary matrices from the diagonalisation drop out.
We then simply have to sum over all mass eigenstates the results for these diagrams computed in Ref.~\cite{Martin:1996zb}.

The next diagram \ref{twoloop7} is slightly more involved.
This diagram arises from the D-term interactions. D-terms distinguish between chiral and antichiral fields,
in our case $\rho,Z$ and $\tilde{\rho},\tilde{Z}$,
respectively. We have defined our scalar field $S$ in \eqref{Sdef} such that all component fields have equal charges.
Accordingly, the ordinary gauge vertex is
proportional to a unit matrix in the component space (cf. Eq. \eqref{gaugeint}).
This vertex is then `dressed' with our diagonalisation matrices when we switch to
the $\hat{S}$ basis, \eqref{gaugeint2}.
This is different for diagram \ref{twoloop7}. Here we have an additional minus-sign between chiral and antichiral fields.
In field space this corresponds to
 a vertex that is proportional to a matrix $V_{D}={\rm diag}(1,1,-1,-1)$.
We therefore obtain,
\begin{equation}
{\rm Fig.}~\ref{twoloop7}\, = \, \sum_{i,m} (Q^{T}V_{D}Q)_{i,m}J(\hat{m}_{0,m},\hat{m}_{0,i})(Q^{T}V_{D}Q)_{m,i},
\end{equation}
where $J$ is the appropriate two-loop integral for Fig.~\ref{twoloop7} which can be found in \cite{Martin:1996zb}.

Finally, in \ref{twoloop8} we have a mixed boson/fermion loop.
The subdiagram containing the messengers is similar to the diagram for the gaugino mass. The only difference is the direction of the
arrows on the gaugino lines. Indeed the one-loop sub-diagram corresponds to a contribution
to the kinetic term rather than a mass term for the gauginos.
(The mass term will of course contribute as well but will be suppressed
by quark masses.)
Using Eq. \eqref{gaugeint2} we find,
\begin{eqnarray}
{\rm Fig.}~ \ref{twoloop8} & = & \sum_{ik}(|U_{i1}^{\dagger}Q_{1k}+U_{i2}^{\dagger}Q_{2k}|^{2}
+|Q_{k3}^{\dagger}V_{1i}+Q_{k4}^{\dagger}V_{2i}|^{2})L(\hat{m}_{1/2,i},\hat{m}_{0,k}^{2})\,,
\end{eqnarray}
where $L$ is again the appropriate loop integral from \cite{Martin:1996zb}.

Summing over all diagrams we find the sfermion masses depicted in Fig. \ref{sfermionmassnum}.
Comparing to the gaugino masses \ref{gauginomassnum} we find the sfermion masses to be significantly bigger.
Indeed, the scalar masses roughly follow the naive estimate
\begin{equation}
m^{2, {\rm naive\,\, estimate}}_{\tilde{f}}\sim \frac{g^4}{(16\pi^2)^{2}} \hat{\mu}^{2}.
\end{equation}
This demonstrates again the fundamental difference between the generation of gaugino masses and the generation of sfermion masses.

The main results of this section, Eqs.~ \eqref{gauginomass} and \eqref{sfermionmass}
give the gaugino and scalar masses generated at the messenger mass scale $\mu$.
It is useful to factor out the particle-type-dependent overall constants and
define the \emph{universal} fermion and scalar mass contributions $m_{1/2}$
and $m_0$, via
\bea
\label{mhalfdef}
m_{\lambda_A}(\mu) := \frac{g_A^2}{16\pi^2}\,\,m_{1/2}\\
\label{mzerodef}
m^{2}_{\tilde{f}}(\mu):= \sum_{A} \frac{g^{4}_{A}}{(16\pi^2)^2}
C_{A} S_{A}\,\,m^2_0
\eea
The main results of this section (2.20) and (2.24) are then expressed in terms of $m_{1/2}$
and $m_0$ which we calculate numerically using the VEVs generated by {\em{Vscape}}.
As an example, in
Table~\ref{masstable} we show the values for $m_{1/2}$ and $m_0$ obtained for the same
parameters as in Table~\ref{vscapetable}. For a more up to date discussion of gaugino and scalar mass values
we refer the reader to a more recent paper \cite{AJKM}. In section 4 of that reference we have used
more accurate numerical values for stabilised VEVs
(with no tree-level constraints imposed) and have also included contributions to $m_{1/2}$ and $m_0$ from
loops of $X$ messenger fields.

\begin{table}[h] \begin{center} \begin{tabular}{|c|c|c|c|} \hline $\mu$ & $m$ & $m_{1/2}$ & $m_0^2$ \\
\hline \hline 10 & 0.3 & 1.03984{\scriptsize$ \times10$}$^{-7}$ & 0.026787 \\ \hline 1.1 & 0.3 & 0.017843 & 4.89783 \\
\hline 1.01 & 0.3 & 0.044771 & 5.12698 \\ \hline 1.003 & 0.3 & 0.052320 & 4.74031 \\ \hline \hline
\end{tabular} \end{center}
\caption{Gaugino mass and sfermion mass-squared coefficients for various parameter points.
All values are in units of $\hat{\mu}$. \label{masstable}}
\end{table}

\subsection{Renormalisation group running, mass spectrum and electroweak symmetry breaking}

In the previous section we calculated the soft SUSY-breaking masses for gaugionos and sfermions at the messenger scale $\mu$.
The Higgs masses $m_{H_1}^2$ and $m_{H_2}^2$ are calculated in the same way as
the sfermion masses above\footnote{We use the GUT normalisation convention for the $g_1$ gauge couplings.}:
\be
m_{H_1}^2(\mu) =
m_{H_2}^2(\mu)=\left(\frac{3}{4}\frac{g^{4}_{2}}{(16\pi^2)^2}+\frac{3}{20}
\frac{g^{4}_{1}}{(16\pi^2)^2} \right)\, m^2_0\\
\ee
The other soft SUSY-breaking terms in the MSSM, such as the $A$-terms and the
$B$-term are generated
at two-loop level. Indeed the diagrams giving rise to the $B$-term require an insertion of the Peccei-Quinn
violating parameter $\mus$ {\em and} a SUSY breaking gaugino ``mass loop". Thus its magnitude
at the messenger scale $\mu$ is of order \cite{rattazzi-sarid}
\be
B\mus \sim  \frac{g^2}{16\pi^2}  m_{\lambda} \mus \sim \frac{g^4}{(16\pi^2)^2}  m_{1/2} \mus  ,
\ee
and is loop suppressed with respect to gaugino masses. For the accuracy required here,
it will be sufficient to take
$
B = 0
$
at the messenger scale.

We now turn to the phenomenology in full, beginning with the SUSY
breaking in the visible sector. The next step is to use the renormalisation group running to determine
the soft SUSY breaking parameters at the weak scale, solve for electroweak symmetry breaking and find the mass spectrum of the MSSM.
We will throughout be using the conventions
of Refs.\cite{Abel:2000vs,softsusy} (with the obvious replacement
$\mu\rightarrow\mus$). The pattern of SUSY breaking here is
expected to be different from the standard gauge mediation form for two reasons.
Firstly our model naturally predicts significantly large values of $m_0$ relative to $m_{1/2}.$
Secondly, for reasons explained above, we take $B=0$
at the messenger scale. The phenomenology of the $B=0$ case has been discussed
in refs.\cite{Babu:1996jf,Dimopoulos:1996yq,Bagger:1996ei,rattazzi-sarid}.
The main prediction is that high $\tan\beta$
is required for the electroweak symmetry breaking.

In order to see why, consider the tree level minimization conditions
in $H_{2}$ and $H_{1}$, which are
\begin{eqnarray}
\label{min-one}
\mus^{2} & = & \frac{m_{H_{1}}^{2}-m_{H_{2}}^{2}\tan^{2}\beta}{\tan^{2}\beta-1}-\frac{m_{Z}^{2}}{2}
 \\
\label{min-two}
B \mus & = & \frac{\sin(2\beta)}{2}(m_{H_{1}}^{2}+m_{H_{2}}^{2}+2\mus^{2}).
\end{eqnarray}
Since $B \mus$ is generated only radiatively, the RHS of the
second equation has to be suppressed by small $\sin(2\beta)$ with
$\beta$ approaching $\pi/2$. One additional feature of the $B$-parameter that complicates the analysis somewhat,
is that as noted in Ref.~\cite{rattazzi-sarid} there is an accidental cancellation of renormalization group
contributions to its running close to the weak scale.
Of course this model becomes more fine-tuned as $m_0\gg m_{1/2}$ since we are after all decoupling the supersymmetry in that limit. It is
worth understanding what has to be fine-tuned. Since $\tan\beta \gg 1$ when $m_0\gg m_{1/2}$ the first equation tells us that we must have
$\mus^2 \approx -m^2_{H_2}$. In order to have a hope of satisfying the second equation there has to be a cancellation
of the terms inside the bracket, $m_{H_{1}}^{2}+m_{H_{2}}^{2}+2\mus^{2} \approx 0$ and therefore $m_{H_1}^2\approx m_{H_2}^2 $
near the minimization scale. This is consistent with large $\tan\beta$ where the top and bottom Yukawa couplings become
approximately degenerate\footnote{ Using the conventional definition \cite{fine-tuning}, the fine-tuning is then of order $\mus/m_Z$.}.

To calculate the spectrum of these models we have modified the
{\em Softsusy2.0} program of Ref. \cite{softsusy}. In its unmodified form this program
finds Yukawa couplings consistent with soft SUSY breaking terms (specified at
the messenger scale $\qmess=\mu$) and electroweak symmetry breaking conditions
(imposed at a scale $\qsusy$ to be discussed later). It is usual to take the
ratio of Higgs vevs $\frac{v_u}{v_d}\equiv\tb$ at $\qsusy$ as an input
parameter instead of the soft SUSY breaking term $B \mus$ at $\qmess$. This term,
and
the SUSY preserving $\mus$ are subsequently determined through the EWSB
conditions \eqref{min-one},\eqref{min-two}.

As the models we are considering have $B \mus = 0$ at $\qmess$ to two loops, $\tb$
is not a free parameter, and must (as noted above) be adjusted in
{\em Softsusy2.0}, so that this boundary condition is met.
In detail the iteration procedure works as follows: initially, a
high value of $\tb$ is chosen and all the gauge and Yukawa couplings are evolved
to $\qmess$. The soft parameters are
then set, as per the SUSY breaking model, including the condition
$B \mus = 0$. The whole system is then evolved down to $\qsusy$,
where $\tan\beta$ is adjusted to bring the program closer to a solution of the EWSB
condition in Eq.(\ref{min-two})  (including the
1-loop corrections to the soft masses $m_{H_1}^2$ and $m_{H_2}^2$ and the
self-energy contributions to the $\overline{DR}$ mass-squared of the axial
Higgs, $m_A^2$ ).
We then run back up to $\qmess$ where we reimpose the soft-breaking boundary conditions,
and whole the process is repeated until the value of $\tan\beta $ converges.

The scale $\qsusy$ at which the tree-level minimisation conditions
\eqref{min-one}, \eqref{min-two} are imposed is chosen so as to minimise the
radiative corrections to the results. It is usually taken to
be $\qsusy \equiv x \sqrt{m_{\tilde t_1}m_{\tilde t_2}}$ where $x$ ({\tt QEWSB} in the
language of Ref.\cite{softsusy}) is a number of order unity.
As we see from Table~\ref{varyqsusy}, the lightest Higgs mass (in the model with
spontaneously broken $R$-symmetry) depends less on scale for lower values of
$\qsusy$, and so in this model we will therefore be using $\qsusy =
0.8\times\sqrt{m_{\tilde t_1}m_{\tilde t_2}}$. Note that only the Higgs masses are
sensitive to this choice and the other parameters are largely unaffected.

What the model we are discussing is predicting in most of its parameter space (i.e. generic $\mu > {\hat{\mu}}$)
is clearly split-SUSY like because of the
suppression of $R$-symmetry violating operators (i.e. $m_{1/2}\ll m_0$ in Table 2). It provides a
first-principles model which can implement split-SUSY \cite{split1,split2}. (For other realisations of split-SUSY
scenarios see e.g. \cite{Langacker:2007ac}.)
Our purpose here however is to examine how close models with radiative $R$-symmetry breaking
can get to the usual gauge mediation scenarios \cite{GR}. For this reason we want to
reduce the $m_0/m_{1/2}$ ratio as far as possible and to take $\mu$ approaching $\hat{\mu}$, i.e. the
last two rows in Table 2. In Table~\ref{spectratable} in Section {\bf 4}, we present a benchmark point (Benchmark Point A)
with the full spectrum of the model for $\mu=1.003 {\hat{\mu}}$, $m=0.3{\hat{\mu}}$. This point
corresponds to a phenomenologically viable region of  parameter space near the
boundary, that has heavy scalars, and light charginos and neutralinos and exhibits electroweak symmetry breaking.
This point is still quite distinct from the usual gauge mediation scenarios, and as we will see in the
next section from predictions of gauge mediation models with {\em explicit} $R$-symmetry breaking \cite{MN,AS}.

\begin{table}
\begin{center}
\begin{tabular}{|c|c|c|c|c|c|c|c|c|}
\hline
$\qsusy$ & $\times\,0.75$ & $\times\,0.80$ & $\times\,0.85$ & $\times\,0.9$ &
$\times\,0.95$ &$\times\,0.99$ & $\times\,1.0$ &
$\times1.01$ \\
\hline
$h_0$ & 124.5 & 124.5 & 124.2 & 124.1 & 123.8 & 101.5 & 93.3 & 78.6 \\
\hline
\end{tabular}
\end{center}
\caption{
Checking the scale dependence of the lightest Higgs mass (in GeV).
\label{varyqsusy}}
\end{table}

\section{Gauge mediation with explicit $R$-breaking}

Here we consider the gauge mediation models of Refs.~\cite{MN,AS}
which are working models of metastable SUSY breaking with a messenger
sector that, as already noted, explicitly breaks the $R$-symmetry
of the ISS sector. The general philosophy is to appeal to the details
of the couplings of the messengers to the electric ISS theory to explain
why the explicit $R$-symmetry breaking is so weak in the effective
theory. The nett result is that one only breaks the $R$-symmetry
by operators suppressed by powers of $M_{Pl}$.

Although the phenomenology is expected to broadly follow that of the
gauge mediation paradigm \cite{GR}, there is a difference. We will
argue that, in the present context the Higgs bilinear $B$ parameter
of the MSSM (the SUSY breaking counterpart of $\mus H_{u}H_{d}$)
is naturally zero at the mediation scale. This is because $R$-symmetry
breaking operators are (by assertion) suppressed by powers of $M_{Pl}$
and this restricts the possibilities for generating the $B$ parameter:
it is either many orders of magnitude too large or forbidden by symmetries
to be zero.

Let us begin by recapping Ref.~\cite{MN} and considering this issue
in detail, before presenting the SUSY breaking phenomenology, and
an example
benchmark point. The model augments the original
ISS model with a pair of messengers {}``quarks'' charged under the
SM gauge group denoted $f$ and $\tilde{f}$ of mass $M_{f}$. For
simplicity we shall assume that they form a fundamental and antifundamental
respectively of the parent $SU(5)$ of the SM. It was proposed that
these couple maximally to the electric theory via a piece of the
form
\begin{equation}
W_{R}=\frac{\lambda}{M_{Pl}}(\tilde{Q}Q)(\tilde{f}f)+M_{f}\tilde{f}f,
\label{WRff}
\end{equation}
where $M_{Pl}$ is the scale of new physics at which the operator
is generated, hereafter assumed to be the Planck scale.
 For simplicity
we shall for this discussion consider both $\mu^{2}$ and $\lambda$
to be flavour independent couplings. The essential observation of
Ref.~\cite{MN} is that, in the magnetic theory, this appears as an
extremely weak violation of $R$-symmetry due to the large energy
scale at which the operator is generated
\begin{equation}
W_{R}=\lambda'\Phi\tilde{f}f+M_{f}\tilde{f}f \equiv S_{mess} \tilde{f}f
\end{equation}
where we introduced spurion superfield $S_{mess}$ as in the standard gauge-mediation set-up.
By assumption the high energy scale $M_{Pl}$ is much larger
than $\Lambda$ so that
\begin{eqnarray}
\lambda'=\frac{\lambda\Lambda}{M_{Pl}} & \ll & 1.
\end{eqnarray}
Since the $R$-symmetry is not respected by $W_{R}$
the Nelson-Seiberg theorem \cite{NS} necessarily leads to the appearance of a new SUSY-preserving vacuum,
but as long as $\lambda'$ is small enough, the transition rate from $\vplus$ to this new vacuum is suppressed and
the original ISS picture is unchanged. Indeed the meson $\Phi$ field can remain trapped in $\vplus$ near the origin,
with effective messenger $F$-term and scalar VEVs of the spurion superfield
\begin{eqnarray}
\langle F_{mess}\rangle & \equiv & \lambda' \langle F_{\Phi}\rangle \, =\, \lambda'\mu^{2}\nonumber \\
\langle S_{mess} \rangle & \equiv &  \lambda' \langle {\Phi} \rangle  + M_{f} \, \approx\,  M_{f}
\end{eqnarray}
As in usual gauge mediation, a gaugino mass is induced at one loop
of order
\begin{equation}
m_{\lambda} \sim \frac{g^{2}}{16\pi^{2}}\frac{\langle F_{mess}\rangle}{\langle S_{mess} \rangle}
\sim\frac{g^{2}}{16\pi^{2}}\frac{\lambda'\mu^{2}}{M_{f}}
\label{eq:gaugino1}
\end{equation}
and a scalar mass-squared of the same order induced at two loops,
 \begin{equation}
m_{\tilde{q}}^2 \sim  m_{\lambda}^2 .
\label{eq:msc1}
\end{equation}
The last equation is a consequence of the fact that $R$-symmetry breaking which controls gaugino masses
is linked to (i.e. not much smaller than) the SUSY-breaking scale of the Visible Sector.

There is a new global minimum where
the rank condition \eqref{rank-cond}
is satisfied and the $\mu^2$-ISS term is cancelled in the ISS potential,
\begin{eqnarray}
\langle\tilde{f}f\rangle & = & \mu^{2}/\lambda'\nonumber \\
\langle\Phi\rangle & = & M_{f}/\lambda',
\end{eqnarray}
however for small enough $\lambda'$ these minima can be much further
from the origin than $\Lambda$, beyond which all that one can say
is there will be a global minimum of order $\langle\Phi\rangle\sim\Lambda$.
Such far-flung minima do not change the ISS picture of metastability,
and this is why the weakness of $\lambda'$ is welcome. The resulting
bound is $M_{f}\gtrsim\lambda'\mu$ \cite{MN}. Coupled with the gaugino
mass being of order $m_{W}$, we find only very weak bounds:
\begin{equation}
\mu\gtrsim16\pi^{2}m_{W}.\label{eq:weakbound}
\end{equation}
 There are a number of additional constraints, two of the most important
being that $f$ is non-tachyonic which gives $M_{f}^{2}>\lambda'\mu^{2}$,
and that gauge mediation is dominant $\frac{M_{f}}{M_{Pl}}\lesssim10^{-4}\lambda'$.
Additional constraints come from the possibility of additional operators
such as $\delta W=\Phi^{2}/M_{Pl}$ which are now allowed in the superpotential,
however all of these can be easily satisfied for high values of $\Lambda$.

\subsection{Forbidden operators and $B=0$}

If one considers the MSSM sector as well, then there are other Planck-suppressed
$R$-symmetry breaking operators that had to be forbidden
in Refs.~\cite{MN,AS}. Normally in gauge mediation one is justified
in neglecting gravitationally induced operators altogether, however as we have seen,
in these models the leading Planck-suppressed operator plays a pivotal
role. Hence it is important to determine what effect other Planck-suppressed operators may have.
The most important conclusion of this
discussion will be that phenomenological consistency requires $B\approx0$
at the mediation scale.

Before considering the operators in question, it is worth recalling
the problem with $B$ in usual gauge mediation, in which supersymmetry
breaking is described by a Hidden sector spurion superfield $S_{mess}$. The problem arises when
one tries to generate the $\mus H_{u}H_{d}$ term of the MSSM
(for a recent review see Ref.~\cite{slavich}). Consider generating
$\mus$ directly in the superpotential. There are two possibilities,
either the parameter $\mus$ depends on $\langle S_{mess}\rangle $ in which
case a $B$-term is generated, or it does not, in which case $B=0$.
Let us suppose that it does, and that the superpotential contains
$W\supset\mus(S_{mess})H_{u}H_{d}$. The $B$ term is given
by
\begin{equation}
B=\frac{\mus'}{\mus}F_{mess}\sim \frac{F_{mess}}{S_{mess}},
\end{equation}
where $\mus' = d \mus/d S_{mess}$ and the final relation follows from a dimensional analysis.
This should be compared with the SUSY breaking contribution to the
gaugino masses which appear at one loop, $m_{\lambda}\sim\frac{g^{2}}{16\pi^{2}}\frac{F_{mess}}{S_{mess}}$,
so that,\begin{equation}
B\sim\frac{(16\pi^{2})}{g^{2}}m_{\lambda}.\end{equation}
Hence one finds that $B \mus$ is two orders of magnitude too
large. More generally because the $\mus$ and $B \mus$
terms are both forbidden by a Peccei-Quinn symmetry, they tend to
be generated at the same order, whereas $B \mus$ should have
an additional loop suppression (in order to be comparable to the scalar
mass-squareds). One can then assume that $\mus$ is independent
of $S_{mess}$ in which case $B=0$, or try to find a more sophisticated
dynamical reason that the $B$ term receives loop suppression factors.

Now let us turn to the models of Ref.~\cite{MN,AS}. Here the situation
is rather more pronounced for the very same reason that the $R$-symmetry
breaking is under control, namely that the spurion is related to a
meson of the electric theory. The $\mus$ term will be a function
of
\be
\frac{Q\tilde{Q}}{M_{Pl}}=\frac{\Lambda\Phi}{M_{Pl}},
\ee
and will be dominated by the leading terms. The leading operators
involving $H_{2}H_{1}$ we can consider are
\begin{eqnarray}
W & \supset & \mu_{0}H_{2}H_{1}+\frac{\lambda_{2}}{M_{Pl}}H_{2}H_{1}\tilde{f}f
+\frac{\lambda_{3}}{M_{Pl}}H_{2}H_{1}\tilde{Q}Q,\nonumber \\
K & \supset & \lambda_{4}\frac{(Q\tilde{Q})^{\dagger}H_{2}H_{1}}{M_{Pl}^{2}}+h.c..
\end{eqnarray}
where $\lambda_{2,3,4}\sim1$. We will for generality allow the $\mu_{0}$
term which is consistent with $R$-symmetry in the renormalizable
theory; this represents supersymmetric contributions to the $\mus$-term
that do not involve the ISS sector. (It would of course be inconsistent
to allow further SUSY breaking in the non-ISS sector.) The remaining
$R$-violating operators we will take to be Planck suppressed as prescribed
in refs.\cite{MN,AS}.

Unfortunately it is clear that the Kahler potential term cannot (as
it could in Refs.~\cite{GuidiceMasiero,rattazzi}) be responsible for
the $\mus$-term. Its contribution is of order \begin{equation}
\mus\sim\frac{\Lambda}{M_{Pl}^{2}}\mu^{2}\end{equation}
but we require $\mu^{2}\ll M_{Pl}m_{W}$ for gauge-mediation to be
dominant, which would imply $\mus\ll \frac{\Lambda}{M_{Pl}} m_{W}$. Similar consideration
apply to operators in the Kahler potential with factors of $D^{2}(\Phi^{\dagger}\Phi)$
as in \cite{dvali}.

Turning instead to the leading superpotential terms, and assuming
the messengers remain VEVless, one has
\begin{eqnarray}
\mus & = & \mu_{0}+\lambda_{3}\frac{\Lambda}{M_{Pl}}
\langle\Phi\rangle\sim\mu_{0}+\lambda_{3}16\pi^{2}\frac{\Lambda^{3}}{M_{Pl}^{2}}\nonumber \\
B \mus & = & \lambda_{3}\frac{\Lambda\mu^{2}}{M_{Pl}}\nonumber \\
m_{Higgs}^{2} & \sim & \frac{g^{4}}{(16\pi^{2})^{2}}\frac{\Lambda^{2}}{M_{Pl}^{2}M_{f}^{2}}\mu^{4}\nonumber \\
m_{\lambda} & \sim & \frac{g^{2}}{16\pi^{2}}\frac{\Lambda}{M_{Pl}}\frac{\mu^{2}}{M_{f}},
\end{eqnarray}
 where we used the fact that, as shown in Ref.~\cite{MN}, the $\Phi$
field is expected to get only a small VEV due to the presence of $R$-symmetry
breaking operators, which was estimated to be
\begin{equation}
\langle\Phi\rangle\sim16\pi^{2}\frac{\Lambda^{2}}{M_{Pl}}.
\end{equation}
The above gives
\begin{equation}
B \mus\sim\frac{16\pi^{2}}{g^{2}}m_{\lambda}M_{f}.
\end{equation}
Typically $M_{f}$ has to be orders of magnitude above $m_{W}$, so
the situation is considerably worse than in usual gauge mediation
unless a symmetry forbids the $\lambda_{3}$ coupling. A global $R$-symmetry
would not be respected by gravitationally suppressed operators, however
it \emph{is} possible that particular operators can be suppressed.
If $\mus$ for example is charged under an additional gauge
symmetry then one might expect
\begin{equation}
\lambda_{2}\sim\lambda_{3}\sim\frac{\mus}{M_{Pl}},
\end{equation}
in which case the effect of these operators is utterly negligible
and we effectively have
\begin{eqnarray}
\mus & \approx & \mu_{0}\nonumber \\
B & \approx & 0.
\end{eqnarray}

Note the importance of the interpretation of the effective ISS theory
as a magnetic dual in this discussion. For example one could also
have considered the effective operator
\begin{equation}
W_{\sst R/MSSM}=\frac{\lambda_{4}}{M_{Pl}}H_{2}H_{1}Tr(\tilde{\varphi}.\varphi).
\end{equation}
This would have given $\mus\sim\frac{\mu^{2}}{M_{Pl}}$ similar
to the Giudice-Masiero mechanism, above. However because the magnetic
quarks $\varphi$ and $\tilde{\varphi}$ are composite objects, the
coupling $\lambda_{4}$ will be suppressed by many powers of $\Lambda/M_{Pl}$
so this contribution to $\mus$ would always be negligible.

Finally,
we have investigated
the pattern of SUSY breaking in the model with explicit $R$-breaking \cite{MN,AS}, discussed above.
As expected,
it conforms to the standard gauge mediation form, with a requirement that $B=0$
at the mediation scale.
In Table~\ref{spectratable} in Section {\bf 4}, we present a benchmark point (Benchmark Point B)
with the full spectrum of the model.

\begin{table}
\begin{center}

\begin{tabular}{|c|c|c|}
\hline
$\quad\quad\quad\quad$& $\,\,$Model A$\,\,$ & $\,\,$Model B$\,\,$
\\
\hline
$\qmess$ & 8.32\tten$^5$ & 1\tten$^7$ \\
\hline
$\tan\beta$ & 58.7 & 38.9 \\
\hline
${\rm sgn} \mus$ & + & + \\
\hline
$\mus(\qsusy)$ & 2891 & 939 \\
\hline
\hline
$\tilde{e}_L, \tilde{\mu}_L$ & 4165 & 747.9 \\
\hline
$\tilde{e}_R, \tilde{\mu}_R$ & 2133 & 399.8  \\
\hline
$\tilde{\tau}_L$ & 1818 & 319.4 \\
\hline
$\tilde{\tau}_R$ & 4093 & 737.5 \\
\hline
$\tilde{u}_1, \tilde{c}_1$ & 11757 & 1963 \\
\hline
$\tilde{u}_2, \tilde{c}_2$ & 11205 & 1867 \\
\hline
$\tilde{t}_1$ & 10345 & 1593 \\
\hline
$\tilde{t}_2$ & 11061 & 1825 \\
\hline
$\tilde{d}_1, \tilde{s}_1$ & 11784 & 1973 \\
\hline
$\tilde{d}_2, \tilde{s}_2$ & 11144 & 1851 \\
\hline
$\tilde{b}_1$ & 10298 & 1754 \\
\hline
$\tilde{b}_2$ & 11060 & 1822 \\
\hline
$\chi_1^0$ & 60.8 & 270.3 \\
\hline
$\chi_2^0$ & 125.0 & 524.8 \\
\hline
$\chi_3^0$ & 2906 & 949.0 \\
\hline
$\chi_4^0$ & 2929 & 950.3 \\
\hline
$\chi_1^{\pm}$ & 100.7 & 526.5 \\
\hline
$\chi_2^{\pm}$ & 2894 & 945.6 \\
\hline
$h_0$ & 124.8 & 137.6 \\
\hline
$A_0$, $H_0$ & 184.5 & 975.1 \\
\hline
$H^{\pm}$ & 207.4 & 978.6 \\
\hline
$\tilde{g}$ & 414.2 & 1500 \\
\hline
$\tilde{\nu}_{1,2}$ & 4175 & 740.2 \\
\hline
$\tilde{\nu}_3$ & 4095 & 724.4 \\
\hline
\hline
\end{tabular}
\end{center}
\caption{
Sparticle spectra for SUSY breaking models with spontaneously broken
(Model A)
and explicitly broken (Model B) $R$-symmetry. All masses are in GeV.
\label{spectratable}}
\end{table}

\section{Conclusions}

It can be argued that in generic models of low scale supersymmetry breaking (where gravity effects can be neglected) metastability
is inevitable.

In this paper we compared SUSY-breaking patterns generated in two distinct and complementary scenarios
of gauge-mediated supersymmetry breaking. Both scenarios employ an explicit formulation of the Hidden Sector
in terms of an ISS-like gauge theory with a long-lived metastable vacuum. This, in both cases, provides a
simple and calculable model to implement metastable DSB.

The difference between the two approaches
lies in the mechanism of $R$-symmetry breaking. The first one, described in Section {\bf 2},
employs spontaneous $R$-symmetry breaking induced by radiative corrections.
It is based on the direct gauge mediation model introduced in \cite{ADJK}.
We find that $R$-symmetry violating soft terms (such as gaugino masses) tend to be suppressed
with respect to $R$-symmetry preserving ones, leading to a scenario with
large scalar masses. These models effectively interpolate between split SUSY models and standard gauge mediation.

The second approach, outlined in Section {\bf 3} is based on gauge mediation models of Refs.~\cite{MN,AS}
with a messenger
sector that explicitly breaks the $R$-symmetry
of the ISS sector
by operators suppressed by powers of $M_{Pl}$.
We argue that these models lead to phenomenology
broadly similar to
standard gauge mediation, but with an additional constraint that $B=0$ at the mediation scale.

Determining the complete spectrum of superpartner masses at benchmark points (see Table~\ref{spectratable})
we find that apart from high values of $\tan \beta$ (arising from the condition that $B \approx 0$
at the messenger scale in both models) the phenomenology of these models is quite different.
For the model with explicit $R$-symmetry breaking (Benchmark Point B) we find that it follows closely the usual gauge mediation scenario
where gauginos and sfermions have roughly equal masses.
In contrast the model with spontaneous $R$-symmetry breaking typically has sfermions that are considerably heavier than the gauginos
-- resembling a scenario of split SUSY.
Benchmark Point A represents such a model at a region in parameter space where the `split aspects' of supersymmetry are
minimal.\footnote{A more up to date discussion of such models can be found in \cite{AJKM}.}
 At the same time it is quite distinct from the usual gauge mediation scenarios, having
relatively heavy scalars and light charginos and neutralinos.

We conclude that details of the dynamics of the Hidden Sector -- the nature of $R$-symmetry and SUSY-breaking --
leave a clear imprint on the phenomenology of the MSSM. Although the
general gauge mediation scenario incorporates both of these
scenarios, there is enough flexibility in GMSB to distinguish them. It
would be interesting to broaden this study to other models with either
spontaneous or explicit $R$-symmetry breaking, to see if the
general pattern outlined here persists.

\section*{Acknowledgements}

We would like to thank Ben Allanach, Steve Martin and Nathan Seiberg for useful comments
and discussions.
CD is supported by an STFC studentship.

\begin{appendix}
\section{The R-symmetry of the baryon-deformed ISS model}

It is known that the $R$-symmetry of the ISS SQCD manifests itself only as an approximate symmetry
of the magnetic formulation which is broken explicitly in the electric theory by the mass terms of electric quarks $m_Q$.
Here we want to quantify this statement and show that the $R$-symmetry breaking in the microscopic theory is controlled
by small parameter, $m_Q /\Lambda = \mu^2 /\Lambda^2 \ll 1$. As such the intrinsic $R$-breaking effects
and deformations can be neglected. This justifies the approach we follow in section {\bf 2} where the $R$-symmetry
of the magnetic theory is used to constrain the allowed deformations. Consequently, the
$R$-symmetry-preserving baryon deformation in Eq.~\eqref{Wbardef} gives a generic superpotential.

We first consider the massless undeformed SQCD theory, its global symmetry is
$SU(N_f)_L \times SU(N_f)_R \times U(1)_B \times U(1)_A \times \overline{U(1)}_R$.
Following the well-established conventions  \cite{Intriligator:1995au} the $\overline{U(1)}_R$ symmetry is
taken to be anomaly-free, and the axial symmetry $U(1)_A$ is anomalous.
(The $U(1)_R$ symmetry of our section {\bf 2} will be constructed below as an anomalous linear combination of the
$\overline{U(1)}_R$, $U(1)_A$ and $U(1)_B$ above.)
Table~\ref{app1table} lists the charges of matter fields of the electric and the magnetic formulations.
\begin{table}[h]
\begin{center}
\begin{tabular}{|c|c|c|c|c|c|}
\hline
&
{\small $SU(N_f)_L$}&
{\small $SU(N_f)_R$}&
{\small $U(1)_{B}$}&
{\small $U(1)_{A}$}&
{\small $\overline{U(1)}_{R}$}\tabularnewline
\hline
\hline
$Q$& $ \square$ & {\bf 1} & 1 & 1 & {\small$\frac{N_f-N_c}{N_f}$} \tabularnewline
$\tilde{Q}$& {\bf 1} & $ \bar\square$ & $-1$ & 1 & {\small$\frac{N_f-N_c}{N_f}$} \tabularnewline
\hline
\hline
$\Lambda$& {\bf 1} &  {\bf 1} & 0 &  {\small$\frac{2N_f}{3N_c-N_f}$} & 0 \tabularnewline
$W$& {\bf 1}& {\bf 1} & 0 & 0 & 2 \tabularnewline
\hline
\hline
$\varphi$& $ \bar\square$ & 1 & {\small$\frac{N_c}{N_f-N_c}$} & {\small$\frac{2N_f-3N_c}{3N_c-N_f}$} & {\small$\frac{N_c}{N_f}$} \tabularnewline
$\tilde{\varphi}$ & 1 & $\square$ & {\small$-\frac{N_c}{N_f-N_c}$} & {\small$\frac{2N_f-3N_c}{3N_c-N_f}$} & {\small$\frac{N_c}{N_f}$} \tabularnewline
$\Phi = \frac{Q\tilde{Q}}{\Lambda}$ & $\square$ & $\bar\square$ & 0 & 2 {\small$-\frac{2N_f}{3N_c-N_f}$}
& {\small$2\frac{N_f-N_c}{N_f}$} \tabularnewline
\hline
\end{tabular}
\end{center}
\caption{Charges under the global $SU(N_f)_L \times SU(N_f)_R \times U(1)_B \times U(1)_A \times \overline{U(1)}_R$
\label{app1table}}
\end{table}
The scale $\Lambda$ is charged only under the $U(1)_A$ which identifies it as the anomalous $U(1)$. In the usual fashion,
the $U(1)_A$-charge of $\Lambda$ in Table~\ref{app1table} is determined from the nonperturbative superpotential, cf. Eq.~\eqref{Wdyn},
\be
W_{\rm dyn}\, =\, (N_f-N_c) \left(  \frac{\det_{\sst \Nf} \tilde{Q}Q}{\Lambda^{3N_c-N_f}}\right)^\frac{1}{N_f-N_c}
\label{Wdyn2}
\ee
Table~\ref{app1table} also shows that the superpotential $W$ is charged only under the $\overline{U(1)}_R$ which
identifies it as the $R$-symmetry (such that $\int d^2 \theta \, W$ is neutral).

Finally, the charges of magnetic quarks $\varphi$, $\tilde{\varphi}$ are derived from the matching between electric and
magnetic baryons, $B_E /\Lambda^{N_c} = B_M /\Lambda^{N_f-N_c}$, $\tilde{B}_E /\Lambda^{N_c} = \tilde{B}_M /\Lambda^{N_f-N_c}$
which implies (schematically)
\be
\left(\frac{\varphi}{\Lambda}\right)^{N_f-N_c} = \left(\frac{Q}{\Lambda}\right)^{N_c} \ , \qquad
\left(\frac{\tilde\varphi}{\Lambda}\right)^{N_f-N_c} = \left(\frac{\tilde{Q}}{\Lambda}\right)^{N_c} \ .
\ee
The charges of $\Phi$ are read off its definition, $\Phi = \frac{Q\tilde{Q}}{\Lambda}$.
As a consistency test on these charges, one can easily verify that the magnetic superpotential $W = \varphi \Phi \tilde\varphi$
is automatically neutral under $U(1)_A$, $U(1)_B$ and has the required charge 2 under the $R$-symmetry.

We now introduce mass terms $m_Q \tilde{Q} Q$ in the superpotential of the electric theory. We want to continue
describing the symmetry structure in terms of the parameters of the IR magnetic theory. For this purpose we use
for quark masses $m_Q =\frac{\mu^2}{\Lambda}$.
This mass-deformation breaks the flavour group $SU(N_f)_L \times SU(N_f)_R $ to the diagonal $SU(N_f)$
(if, for example, all quark masses were the same). It also breaks $U(1)_A \times \overline{U(1)}_R$,
to a linear combination $U(1)$ subgroup. If in addition, we introduce the baryon deformation, as in section
{\bf 2},  it breaks the third $U(1)_B$ factor. In total, the combined effect of the two deformations breaks
 $U(1)_B \times U(1)_A \times \overline{U(1)}_R$ to a single $U(1)_R$. This is the $R$-symmetry of section {\bf 2}
 and it is anomalous since $\Lambda$ is charged under it.\footnote{Note that the two deformations are associated with
 orthogonal $U(1)$'s and are therefore independent.}

 To explicitly construct this surviving $U(1)_R$ for the model of section {\bf 2},
 we set $N_c=5$ and $N_f=7$ and list the three $U(1)$ charges in Table~\ref{app2table}.
 \begin{table}
\begin{center}
\begin{tabular}{|c|c|c|c|}
\hline
&
{\small $U(1)_{B}$}&
{\small $U(1)_{A}$}&
{\small $\overline{U(1)}_{R}$}\tabularnewline
\hline
\hline
$Q$&  1 & 1 & {\small$\frac{2}{7}$} \tabularnewline
$\tilde{Q}$& $-1$ & 1 & {\small$\frac{2}{7}$} \tabularnewline
\hline
\hline
$\Lambda$& 0 & {\small$\frac{7}{4}$} & 0 \tabularnewline
$W$&  0 & 0 & 2 \tabularnewline
\hline
\hline
$\varphi$& {\small$\frac{5}{2}$} &{\small$-\frac{1}{8}$} & {\small$\frac{5}{7}$} \tabularnewline
$\tilde{\varphi}$ & {\small$-\frac{5}{2}$} &{\small$-\frac{1}{8}$} & {\small$\frac{5}{7}$} \tabularnewline
$\Phi$ & 0 &{\small$\frac{1}{4}$} & {\small$\frac{4}{7}$} \tabularnewline
\hline
\end{tabular}
\end{center}
\caption{Charges under  $U(1)_B \times U(1)_A \times \overline{U(1)}_R$ for $N_c=5$ and $N_f=7$
\label{app2table}}
\end{table}
It is now clear that the $U(1)_R$ symmetry of section {\bf 2} is the linear combination
of the three $U(1)$'s with the charge
\be
R \, =\, \overline{R} + \frac{40}{7} \, A + \frac{2}{5} \, B
\label{Rsymour}
\ee
This is the unique unbroken linear combination surviving the mass- plus the baryon-deformation,
$-\mu^2 \Phi + m \varphi^2$, of the magnetic theory
with the charges listed in Table~\ref{app3table}.
 \begin{table}
\begin{center}
\begin{tabular}{|c|c|}
\hline
&
{\small ${U(1)}_{R}$}\tabularnewline
\hline
\hline
$\varphi$& 1 \tabularnewline
$\tilde{\varphi}$ & $-1$ \tabularnewline
$\Phi$ & 2 \tabularnewline
\hline
\hline
$\Lambda$& 10 \tabularnewline
$\mu$ & 0 \tabularnewline
$W$&  2 \tabularnewline
\hline
\hline
$Q$& {\small$6+\frac{2}{5}$} \tabularnewline
$\tilde{Q}$& {\small$6-\frac{2}{5}$} \tabularnewline
\hline
$m_Q=\frac{\mu^2}{\Lambda}$ & $-10$ \tabularnewline
\hline
\end{tabular}
\end{center}
\caption{Charges under ${U(1)}_R$ for $N_c=5$ and $N_f=7$
\label{app3table}}
\end{table}
In the magnetic Seiberg-dual formulation, the $U(1)_R$ symmetry is manifest. It is the symmetry of the
perturbative superpotential \eqref{Wbardef} which is only broken anomalously.

In the electric formulation
the $U(1)_R$ symmetry is broken by the mass terms $m_Q \, \tilde{Q}Q$ on the account of the explicit $\Lambda$-dependence of the masses
 $m_Q =\frac{\mu^2}{\Lambda}$. It is also broken by the baryon deformation (again in the electric theory language)
 $\frac{1}{M_{Pl}^2} \, Q^5$ because the magnetic baryon deformation parameter $m$ in \eqref{mbardef}
 explicitly depends on $\Lambda$.
 Thus the apparent $U(1)_R$ symmetry of the IR theory is only approximate, and
 is lifted in the UV theory. However, the $R$-symmetry is broken in a controlled way, by the parameter of the order
 of $m_Q / \Lambda$.
 To verify this note that in the limit $m_Q \to 0$, the electric quark masses disappear while the baryon deformation
 $\frac{1}{M_{Pl}^2} \, Q^5$ is invariant under the $R$-symmetry $U(1)_{R'}$:
 \be
R' \, =\, \overline{R} + \frac{5}{7} \, A - \frac{3}{5} \, B
\label{Rsymnew}
\ee
This is a different from \eqref{Rsymour} linear combination, but in the massless limit we are considering it is a
perfectly valid classically conserved $R$-symmetry which protects the baryon deformation in the electric theory and forbids
e.g. anti-baryon deformations $\frac{1}{M_{Pl}^2} \, \tilde{Q}^5$. Thus in the massless limit there is always an $R$-symmetry
which protects baryon deformations either in the electric or in the magnetic formulation. When quark masses are non-vanishing,
this $R$-symmetry is broken by $m_Q / \Lambda$.
Indeed, if one formally sends $\Lambda \to \infty$ holding $\mu$ and $m$ fixed,
 the dynamical non-perturbative superpotential disappears
 and the exact $U(1)_R$ is recovered.

 In general, anomalous global symmetries do not match in the magnetic and the electric descriptions.
 The $U(1)_R$ is an approximate symmetry and in principle one should allow generic $U(1)_R$-violating
 deformations. For example, one can add an antibaryon $\tilde{B}$ deformation to the superpotential \eqref{Wbardef}.
 However, these deformations are suppressed relative to the $U(1)_R$-preserving ones
 by the small parameter, $m_Q /\Lambda = \mu^2 /\Lambda^2 \ll 1$, and therefore can be neglected.

\end{appendix}

\end{document}